\documentclass[10pt,tightenlines,aps,prd,twocolumn,showpacs,preprintnumbers,amsmath,amssymb]{revtex4}
\usepackage{graphicx,color}
\definecolor{darkblue}{rgb}{0,0,0.7}
\definecolor{darkred}{rgb}{0.7,0,0}
\usepackage[dvips, colorlinks, citecolor=darkblue, linkcolor=darkred, urlcolor=blue]{hyperref}
\allowdisplaybreaks
\begin{document}

\title{Displacement- and laser-noise-free gravitational-wave detection
with two Fabry-Perot cavities}
\author{Andrey A. Rakhubovsky and Sergey P. Vyatchanin}
\affiliation{Faculty of Physics, Moscow State University, Moscow, 119992, Russia} 
\email{svyatchanin@phys.msu.ru}
\date{\today}

\begin{abstract}
We propose  two Fabry-Perot cavities, each pumped through both the mirrors, positioned in line as \textit{a toy model} of the gravitational-wave (GW) detector free from displacement noise of the test masses. It is demonstrated that the displacement noise of cavity mirrors as well as laser noise can be completely excluded in a proper linear combination of the cavities output signals. We show that in low-frequency approximation (gravitational wave length $\lambda-\text{gw}$ is much greater than distance $L$ between mirrors $\lambda_\text{gw}\gg L$)  the decrease of response signal is about $(L/\lambda_\text{gw})^2$, i.e. signal is stronger than the one of the interferometer recently proposed by S.~Kawamura and Y.~Chen \cite{2004_DNF_GW_detection}.
\end{abstract}

\maketitle

\section{Introduction}\label{sec_intro}
Currently there is an international ``community'' of the first generation laser interferometric gravitational wave (GW) detectors \cite{1972_EMW_detector,2000_GW_detection} (LIGO in
USA \cite{1992_LIGO,2006_LIGO_status,website_LIGO}, VIRGO in Italy
\cite{2006_VIRGO_status,website_VIRGO}, GEO-600 in Germany
\cite{2006_GEO-600_status,website_GEO-600}, TAMA-300 in Japan
\cite{2005_TAMA-300_status,website_TAMA-300} and ACIGA in Australia
\cite{2006_ACIGA_status,website_ACIGA}). The development of the
second-generation GW detectors (Advanced LIGO in USA
\cite{2002_Adv_LIGO_config,website_Adv_LIGO}, LCGT in Japan
\cite{2006_LCGT_status}) is underway.
The ultimate sensitivity of laser GW detectors is restricted by the Standard Quantum Limit (SQL) --- a specific sensitivity level where the measurement noise of the meter (photon shot noise) is equal to its back-action noise (radiation pressure noise) \cite{1968_SQL,1975_SQL,1977_SQL,1992_quant_meas}. The sensitivity of GW detectors is also limited by classical displacements noises of various nature: seismic and
gravity-gradient noise at low frequencies (below $\sim 50$ Hz), thermal noise in suspensions, bulk and coatings of the mirrors ($\sim 50\div 500$ Hz).

In 2004 S. Kawamura and Y. Chen put forward an idea of so called displacement-noise-free
interferometer (DFI) which is free from displacement noise of the test masses as well as from optical laser noise \cite{2004_DNF_GW_detection, 2006_DTNF_GW_detection,2006_interferometers_DNF_GW_detection}.
The most attractive feature of DFI is the ability to achieve sub-SQL sensitivity (no SQL since radiation pressure noise is canceled) not accompanied by the necessity to build complicated optical schemes for Quantum-Non-Demolition (QND) measurements \cite{1981_squeezed_light,1982_squeezed_light,2002_conversion,1996_QND_toys_tools}.

The possibility to separate GW signal from displacement noise of the test masses is based on the {\em distributed} character of interaction between GW and light wave unlike the localized influence of mirrors positions on the light wave, taking place only at the moments of reflection. The ``price'' for this separation is the decreased detector response to GWs, especially at low frequencies where the so called long wave approximation is valid, that is when the distance $L$ separating test masses is much less than the gravitational wave length $\lambda_\text{gw}$, i.e. $L/\lambda_\text{gw}\ll 1$ or $\Omega_{\textrm{gw}}\tau\ll 1$ ($\tau =L/c$ is the light travel time between test masses, $c$ is the speed of light and $\Omega_\text{gw}=2\pi\, c/\lambda_\text{gw}$ is the GW frequency). In particular, the analysis presented in \cite{2006_interferometers_DNF_GW_detection} for double Mach-Zehnder interferometer showed that in long wave approximation the shot-noise limited sensitivity to GWs turns out to be limited by $(\Omega_{\textrm{gw}}\tau)^2$-factor for 3D configurations and $(\Omega_{\textrm{gw}}\tau)^3$-factor for 2D configurations. For signals centered at  $\Omega_{\textrm{gw}}/2\pi\approx 100$ Hz and for interferometers with size of $L\approx 4$ km ($\tau\simeq 10^{-5}$~s), DFI sensitivity of the ground-based detector is $(\Omega_\text{gw}\tau)^3\simeq 10^{-6}$ times worse than that of a conventional single round-trip laser detector.

Another approach to the displacement noise cancellation was presented in \cite{2008_toy} where
a single detuned Fabry-Perot  cavity pumped through both of its movable,
partially transparent mirrors was analyzed.

In this paper we investigate model originated from a simple toy model \cite{2008_toy} of the GW detector. Our model consists of two double pumped Fabry-Perot cavities positioned in line. Each cavity is pumped through both  partially transparent mirrors.   By properly combining the signals of  output ports of the cavity an experimenter can remove the information about the fluctuations of the mirrors displacements and laser noise from the data.  The ``price'' for isolation of the GW signal from displacement noise in our case is the suppression  of sensitivity by factor of $(\Omega_{\textrm{gw}}\tau)^2$ (resonance gain partially compensates it) as compared with conventional interferometers --- it is larger than limiting factor $(\Omega_{\textrm{gw}}L/c)^3$ of the double Mach-Zehnder 2D configuration \cite{2006_interferometers_DNF_GW_detection}. 

This paper is organized as follows. In Sec.~\ref{simple} we analyzed simplified round trip model (without any Fabry-Perot cavities). In Sec. \ref{sec_FP_cavity} we derive the response signals of a single double pumped Fabry-Perot cavity to a gravitational wave of arbitrary frequency and introduce their proper linear combination which cancels the laser noise and the fluctuating displacements of one of the mirrors. In Sec. \ref{sec_2FP_cavities} we analyze configuration of two  double-pumped Fabry-Perot cavities which allows to calcel displacement noise of all mirrors completely. Finally in Sec. \ref{conclusion} we discuss the physical meaning of the obtained results and briefly outline the further prospects.

\section{Simplified round trip model}\label{simple}

\begin{figure}[h,t]
%\psfrag{xa}{$x_a$} \psfrag{xb}{$x_b$} \psfrag{xc}{$x_c$}\psfrag{L}{$L$}
%\psfrag{A}{$A$} \psfrag{B}{$B$} \psfrag{C}{$C$}
%\psfrag{E}{$E$} \psfrag{F}{$F$}
\includegraphics[width=0.35\textwidth]{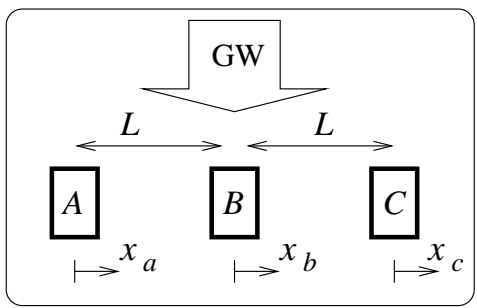}
\caption{Simplified model of displacement noise-free detector. On each platforms we place laser, detectors and reflecting mirrors. Mean distances between neighboring platforms are equal to $L$. GW propagates perpendicularly to line consisting of three platform.}\label{DL3}
\end{figure}

For clear demonstration we start from analysis of the simplest toy model \cite{2004_DNF_GW_detection} consisting of  $3$ platforms $A$, $B$ and $C$ positioned in line as shown on Fig.~\ref{DL3}. GW propagates perpendicularly to this line. We assume that lasers, detectors and mirrors are rigidly mounted on each platform which, in turn, can move as a free masses.  We also assume that mean frequency $\omega_0$ of each laser is equal to  others. In this section we do not take into account laser noise yet paying attention only on displacement noise and GW signal.

We restrict ourselves to the case when radiation emitted from the laser on some platform  is registered (after reflection) by detector on the same platform --- so called round trip configuration.  Actually detectors are homodyne detectors measuring the phase of incident wave.
 
Strictly speaking, in order to describe detection of light wave we have to work in the reference frame of detector, i.e. in accelerated frame. However, in our model detector is mounted on the same platform as laser which radiation detector registers and we can work in inertial laboratory frame as it was demonstrated in  \cite{2008_toy,2008_Tarabrin}. Moreover, in this case of round trip configuration we can use transverse-traceless (TT) gauge  considering GW action as effective modulation of refractive index $(1+h(t)/2)$ by weak GW perturbation metric $h(t)$. It is worth noting that in the opposite case, when laser and detector are mounted on different platforms, we should use the local Lorentz (LL) gauge --- see details in \cite{2008_Tarabrin}.

We denote the phase of the wave  emitted, for example, from platform $A$, reflected on platform $C$ and detected on platform $A$ as $\phi_{aca}$ and so on.  Let us measure  phase $\phi_{aba}$ (of the wave emitted from  and detected on platform $A$ after reflection from platform $B$) and phase $\phi_{bab}$ (see also Fig.~\ref{DL3})
\begin{align}
 %\phi_{ab} &= \psi_h +\Delta_h+ k\big[\tilde x_b^\tau-x_a^\tau\big], \\
 \label{phi_aba}
 \phi_{aba}(t) &= \psi_h(t)+   k\big[2x_b(t-\tau) -x_a(t) - x_a(t-2\tau)\big],\\
 \label{phi_bab}
 \phi_{bab}(t) &= \psi_h(t)+   k\big[-2x_a(t-\tau) +x_b(t) + x_b(t-2\tau)\big],\\
 \label{psi}
 \psi_h (t) & \equiv\frac{\omega_0}{2}\int_{t-2\tau}^{t}h(t')dt',
\end{align}
Here $k=\omega_0/c$ is the wave vector of light emitted by laser, $\tau=L/c$ is bouncing time and $h(t)$ is perturbation of dimensionless metric originated by GW, $c$ is the speed of light.

Obviously, we can exclude information on displacement $x_a$ of platform $A$ in the following combination $\tilde C_1$:
\begin{align}
 \tilde C_1 (t)&= 2\phi_{aba} - \phi_{bab}(t+\tau)-\phi_{bab}(t-\tau)=\nonumber\\
 \label{CC1}
   &= 2\psi_h(t) - \psi_h(t+\tau) -\psi_h(t-\tau)+ \\ 
   &\qquad +k\big[2x_b(t-\tau) -x_b(t+\tau)- x_b(t-3\tau)\big].\nonumber
\end{align}
Exclusion of information on displacements of platforms $A$ in combination $\tilde C_1$ means that we effectively convert platform $A$  into ideal (i.e. displacement noise free) test mass for GW detection.

By similar way measuring phases $\phi_{bcb}$ and $\phi_{cbc}$
\begin{align}
 %\phi_{ab} &= \psi_h +\Delta_h+ k\big[\tilde x_b^\tau-x_a^\tau\big], \\
 \phi_{cbc}(t) &= \psi_h(t)+   k\big[-2x_b(t-\tau) +x_c(t) + x_c(t-2\tau)\big],\nonumber\\
 \phi_{bcb}(t) &= \psi_h(t)+   k\big[2x_c(t-\tau) -x_b(t) - x_b(t-2\tau)\big],\nonumber
\end{align}
we can exclude information on displacement $x_c$ of platform $C$ in combination $\tilde C_2$:
 \begin{align}
  \tilde C_2 (t) &= 2\phi_{cbc} (t) -\phi_{bcb} (t-\tau)-\phi_{bcb} (t+\tau)=\nonumber\\
  \label{CC2}
    &=2\psi_h(t)-\psi_h(t-\tau) - \psi_h(t+\tau)+\\
 &\qquad +  k\big[-2x_b(t-\tau) +x_b(t+\tau) + x_b(t-3\tau)\big]\nonumber
 \end{align}

Comparing (\ref{CC1}) and (\ref{CC2}) we see that position $x_b$ makes contributions into $\tilde C_1$ and $\tilde C_2$ with opposite signs --- in contrast to the GW signal. So we should just sum $\tilde C_1$ and $\tilde C_2$ in order to exclude {\em completely} information on positions of all platforms:
\begin{align}
 \tilde C_3 (t) &= \frac{\tilde C_1(t)+\tilde C_2(t)}{2}= 2\psi_h(t)-\psi_h(t+\tau)-\psi_h(t-\tau)
\end{align}
It is useful to rewrite this formula in frequency domain:
\begin{align}
\psi_h (\Omega)&= \omega_0\tau h(\Omega)\, e^{i\Omega\tau} \, \frac{\sin\Omega\tau}{\Omega\tau}\\
\label{tildeC3}
 \tilde C_3(\Omega) &=-\omega_0\tau\, h(\Omega)\, \left(1 - e^{i\Omega\tau}\right)^2
    \left(\frac{\sin\Omega\tau}{\Omega\tau}\right)
\end{align}
In long  wave approximation ($\Omega\tau \ll 1$) we have in time and frequency domain correspondingly
\begin{align}
 \tilde C_3 (t)&\simeq -\omega_0\tau^3\, \ddot h(t),\\
 \label{tildeC3app}
 \tilde C_3(\Omega) &\simeq \omega_0\tau\, (\Omega\tau)^2\, h(\Omega)
\end{align}

We see that in our simplest model the payment for separation of GW signal from displacement noise is decrease of GW response, which in long wave approximation is about $(\Omega\tau)^2$.

\section{Response of double pumped Fabry-Perot cavity to a gravitational wave}\label{sec_FP_cavity}

\begin{figure}
\includegraphics[width=0.5\textwidth]{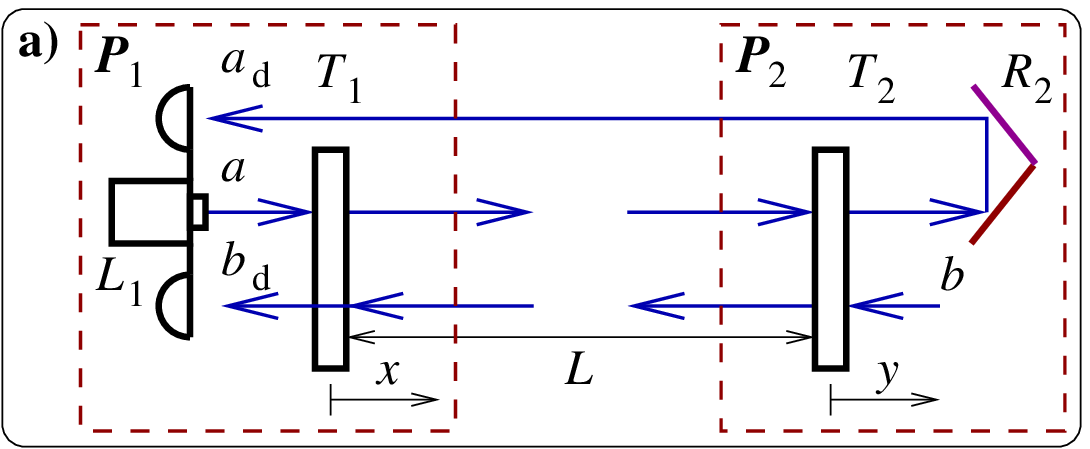}
\includegraphics[width=0.5\textwidth]{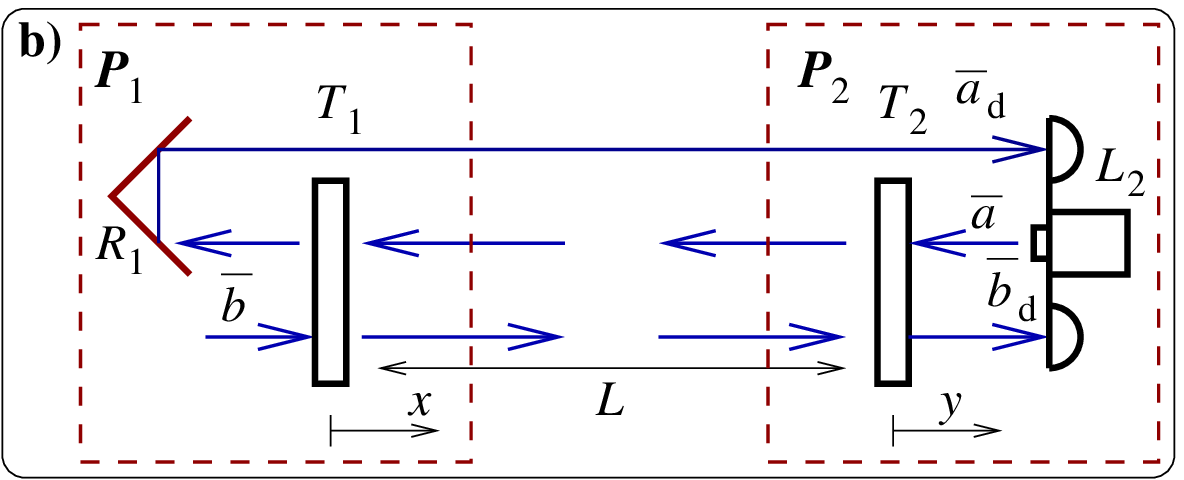}
\caption{Emission-detection scheme of {\em one} double pumped Fabry-Perot cavity. a) pump by laser $L_1$ through the left port is shown only. Pump laser with both detectors and input mirror are assumed to be rigidly mounted on moveable platform $\textrm{P}_1$. Transmitted wave is redirected by additional mirror $R_2$ to platform $P_1$. Transmitted and reflected wave are detected by detectors on platform $P_1$. End and additional mirror $R_2$ are assumed to be rigidly mounted on movable platform $\textrm{P}_2$. b) pump by laser $L_2$ through the right port of the same cavity with its detectors and redirecting mirror $R_1$ is shown.}
\label{T1T2oneFP}
%\end{center}
\end{figure}

Now we can analyze model with two Fabry-Perot cavities. We start from  single double pumped Fabry-Perot cavity presented on Fig.~\ref{T1T2oneFP}. Pump waves in different input ports are assumed to be orthogonally polarized in order the corresponding output waves to be separately detectable and to exclude nonlinear coupling of the corresponding intracavity waves. To simplify our  model we assume that mirrors and lasers with detectors of each cavity are rigidly mounted on two movable platform (see Fig.~\ref{T1T2oneFP}) (in contrast to scheme analyzed in \cite{2008_toy} with four platforms). Laser $L_1$  with its detectors and mirror with amplitude transmittance $T_1$ are rigidly mounted on movable platform $P_1$. In other words, we assume that all the elements on the platform do not move with respect to each other. Laser $L_1$ pumps the cavity from the left and we assume that the wave transmitted through the cavity is redirected to platform $P_1$ by reflecting mirror $R_2$ as shown on Fig~\ref{T1T2oneFP}a. So  waves,  emitted by this laser, are finally registered by detectors positioned on the same platform as laser. The  mirror with amplitude transmittance $T_2$ and laser $L_2$ pumping cavity from the right with its detectors are rigidly mounted on platform $P_2$.  We assume that amplitude transmission coefficients of mirrors are small: $T_1,\ T_2\ll 1$. We put mean distance between the mirrors to be equal to $L$. Without the loss of generality we assume the cavity to be lying in the plane perpendicular to direction of GW and  along one of the GW principal axes.

It is convenient to represent the electric field operator of the light wave as a sum of (i) the ``strong'' (classical) plane monochromatic wave (which approximates the light beam with cross-section $S$) with
amplitude $A$ and frequency $\omega_0$ and (ii) the ``weak'' wave describing quantum fluctuations of the electromagnetic field:
\begin{subequations}
\label{E}
\begin{align}
    E(x,t)&=\sqrt{\frac{2\pi\hbar\omega_0}{Sc}}\,
    \Bigl[A+a(x,t)\Bigr]e^{-i(\omega_0t\mp k_0x)}+{\textrm{h.c.}},\\
    a(x,t)&=\int_{-\infty}^{+\infty}
    a(\omega_0+\Omega)e^{-i\Omega\left(t\mp x/c\right)}\,\frac{d\Omega}{2\pi},\nonumber
\end{align}
\end{subequations}
with amplitude $a(\omega_0+\Omega)$ (Heisenberg operator to be
strict) obeying the commutation relations:
\begin{align*}
    \bigl[a(\omega_0+\Omega),a(\omega_0+\Omega')\bigr]&=0,\\
    \bigl[a(\omega_0+\Omega),a^\dag (\omega_0+\Omega')\bigr]&=2\pi\delta(\Omega-\Omega').
\end{align*}
For briefness throughout the paper we denote
\begin{equation}
a\equiv a(\omega_0+\Omega),\quad a^\dag _-\equiv a^\dag (\omega_0-\Omega) \nonumber
\end{equation}
This notation for quantum fluctuations $a$ is convenient since it coincides exactly with the Fourier representation of the classical fields. 
and we omit the $\sqrt{2\pi\hbar\omega_0/Sc}$-multiplier. For convenience throughout the paper we denote mean amplitudes by block letters and corresponding small additions by {\em the same} small letter as in (\ref{E}). In ideal case the input laser wave is in coherent state (it means that fluctuational amplitude $a(\omega_0+\Omega)$ describes vacuum fluctuations). In more realistic case small amplitudes $a,\ a^\dag $ describes technical laser fluctuations. But fluctuational wave incoming into cavity through the non-pumped port (denoted by $b$ or $\bar b$ on Fig.~\ref{T1T2oneFP}) is always in vacuum state.

In our model, as in simplified model analyzed in previous section, detectors are mounted on the same platform as laser which radiation detectors register and we can work in inertial laboratory frame   \cite{2008_toy,2008_Tarabrin} considering GW action as effective modulation of refractive index $(1+h(\Omega)/2)$ by weak GW perturbation metric $h(\Omega)$.

First, we consider pump by laser $L_1$ shown in the Fig.~\ref{T1T2oneFP}a. Using calculations presented in Appendix \ref{derivation} we can write down formulas for small complex amplitudes $a_d,\ b_d$ of waves detected on platform $P_1$ (see notations on Fig.~\ref{T1T2oneFP}a):
\begin{align}
\label{ad}
a_d&=-\theta_0\psi\big[{\cal T} a+ {\cal R}_2 b \big]+\\
   & + \frac{iT_2\vartheta_0^2}{1-R_1R_2\vartheta_0^2\psi^2}
        \left(\frac{1 +\psi^2}{2}u_x-\psi (u_y+u_h)\right),\nonumber\\
\label{bd}
b_d&={\cal R}_1 a+ {\cal T} b +\\
&   +     \frac{iT_1R_2\vartheta_0^2}{1-R_1R_2\vartheta_0^2\psi^2}
   \left( \frac{1+\psi^2}{2} u_x- \psi (u_y+u_h)\right),\nonumber\\
&\text{where}\quad  \psi = e^{i\Omega\tau},\quad \theta_0=e^{i\delta\tau},\quad \tau=\frac{L}{c}\, .
\end{align}
Here fluctuational amplitudes $a$ and $b$ describe laser noise and vacuum fluctuations correspondingly,
$\delta$ is detuning between laser frequency and resonance frequency of cavity. $R_1=\sqrt{1-T_1^2}$, $R_2=\sqrt{1-T_2^2}$ are reflectivities of mirrors, by calligraph letters we denote coefficients of cavity's transparency and reflectivities:
\begin{align}
 {\cal T} & =\frac{-\vartheta_0\psi T_1T_2}{1-R_1R_2\vartheta_0^2\psi^2},\\
 {\cal R}_1 &= \frac{R_2\vartheta_0^2\psi^2-R_1}{1-R_1R_2\vartheta_0^2\psi^2},\quad
 {\cal R}_2 = \frac{R_1\vartheta_0^2\psi^2-R_2}{1-R_1R_2\vartheta_0^2\psi^2}
\end{align}
The influence of fluctuational (non-geodesic) displacements $x,\ y$ in (\ref{ad}, \ref{bd}) (to be strict its Fourier representations) is described by values $u_x, \ u_y$:
\begin{align}
 u_x &= A_{in} \, 2ik x(\Omega), \quad u_y=A_{in}\, 2iky(\Omega),\\
 \label{Ain}
  A_{in}&=\frac{iT_1A}{1-R_1R_2\vartheta_0^2},\quad
    k=\frac{\omega_0}{c},
\end{align}
where $A_{in}$ is mean amplitude of wave circulating inside the cavity, we assume $A_{in}$ to be real (see also Fig.~\ref{fpthroughT1T2fourP} in Appendix~\ref{derivation}), $A$ is mean amplitude of wave emitted by laser $L_1$ (to be strict amplitude of wave falling on mirror with transparency $T_1$). Interaction of light with GW in (\ref{ad}, \ref{bd}) is described by dimensionless metric perturbation $h$ through value $u_h$:
\begin{align}
 u_h& = A_{in}\, ikL\,h(\Omega)\, \frac{\sin \Omega\tau}{\Omega\tau}\, .
\end{align}

It is worth emphasizing that both output waves $a_d,\ b_d$ contain the identical information on displacements and GW signal --- see formulas (\ref{ad}, \ref{bd}). However, terms describing laser fluctuations have different coefficients at laser noise amplitude $a$. Hence, we can take such linear combination of two detectors output signals which does not contain laser noise (but it will contain the  information on GW signal and displacements). Recall, that in fact we have the homodyne detectors, which can measure arbitrary quadrature component of output waves (with pump laser used as a local oscillator). Our analysis shows that complete cancellation of laser noise is possible at two conditions: i) we should measure the same quadrature in both detector ports; ii) detuning should be zero. For zero detuning only phase quadrature contains information on GW signal and displacements (amplitude quadratures are free from GW signal in linear approximation). Therefore, below we consider the case of detecting phase quadratures at zero detuning. For phase quadrature one can obtain the following formulas  (see details in Appendix~\ref{derivation})
\begin{align}
  a_d^{(p)} &= \frac{T_1T_2\psi^2 \big(a+a^\dag_-\big) -
    \psi(R_1\psi^2-R_2)\big(b+b_{-}^\dag\big)}{\sqrt 2\big(1-R_1R_2\psi^2\big)}+\nonumber\\
\label{adPh}
&\quad         +\frac{2iT_2 }{\sqrt2 \big(1-R_1R_2\psi^2\big)}
        \left(\frac{1+\psi^2}{2}\, u_x - \psi(u_y+u_h)\right);\\
\label{bdPh}
b_d^{(p)}&= \frac{(R_2\psi^2-R_1)\,(a +a^\dag_-)-T_1T_2\psi \, (b+b_{-}^\dag\big)}{
    \sqrt2\big(1-R_1R_2\psi^2\big)}+\\
&\quad +    \frac{2iT_1R_2}{\sqrt2\big(1-R_1R_2\psi^2\big)}\left(
        \frac{1+\psi^2}{2}\, u_x-\psi\, \big(u_y+u_h\big)\right).\nonumber
\end{align}

We see that  laser noise amplitudes contribute to output amplitude quadratures $a_d^{(a)},\ b_d^{(a)}$ in the same combination ($a+a^\dag_-$). Hence, we take linear combinations $C_{ph}=S_a(\Omega) a_d^{(a)} + S_b(\Omega) b_d^{(a)}$ and in order to exclude technical laser noise we specify weight coefficients $S_a,\ S_b$ as following:
\begin{align}
S_a(\Omega)&=\frac{R_2\psi^2-R_1}{1-R_1R_2\psi^2},\quad
    S_b(\Omega)=\frac{-T_1T_2\psi^2}{1-R_1R_2\psi^2},\\
 \label{Cph}
C_{ph}  &= \frac{\big(\psi^2-R_1R_2\big)\psi \, (b+b_{-}^\dag\big)}{
    \sqrt2\big(1-R_1R_2\psi^2\big)}+\\
&\quad +
        \frac{-2iT_2R_1 }{\sqrt2 \big(1-R_1R_2\psi^2\big)}
         \left(\frac{1+\psi^2}{2}\, u_x - \psi(u_y+u_h)\right) .\nonumber    
\end{align}
Here we use normalization $|S_a|^2+|S_b|^2=1$. So we completely cancel laser noise (i.e. combination $C_{ph}$ contains  no term proportional to $(a+ a^\dag_-)$, only vacuum noise $\sim (b+b_-^\dag)$ present).

The dependence of weight coefficients $S_a,\ S_b$ on frequency $\Omega$ mean that before summation output currents of homodyne detectors registering phase quadratures $a_d^{(p)},\ b_d^{(p)}$ should be passed through filters with transmission coefficients $S_a(\Omega),\ S_b(\Omega)$ correspondingly.

Now we can write down formulas for output fields pumping by laser $L_2$  from opposite port (see Fig.~\ref{T1T2oneFP}b). We assume that radiation from laser $L_2$ is polarized normally to radiation emitted by laser $L_1$. We denote all values by the same letters as above but mark them by bar~$\bar{}$. For simplicity we assume that excited by laser $L_2$ mean amplitude $\bar A_{in}$ of the wave circulating inside the cavity is equal to $A_{in}$: $\bar A_{in}=A_{in}$. Also  we  assume that laser $L_2$ is also tuned in resonance (i.e. $\bar \delta=0$) and we measure phase quadratures in corresponding output waves. Again we take corresponding combination $\bar C_{ph}$ to exclude laser noise. Then by using the following substitutions:
\begin{align*}
 T_{1,2}&\to T_{2,1},\quad R_{1,2}\to R_{2,1}, \\
   u_x & \to - u_y,\quad u_y \to -u_x,\quad u_h\to u_h
\end{align*}
we rewrite formula (\ref{Cph}) for combination $\bar C_{ph}$
\begin{align}
 \label{barCph}
\bar C_{ph}  &= \frac{\big(\psi^2-R_1R_2\big)\psi \, (\bar b+\bar b_{-}^\dag\big)}{
    \sqrt2\big(1-R_1R_2\psi^2\big)}+\\
&\quad +
        \frac{-2iT_1R_2 }{\sqrt2 \big(1-R_1R_2\psi^2\big)}
         \left(\frac{1+\psi^2}{2}\, (-u_y) + \psi(u_x+u_h)\right) .\nonumber    
\end{align}

\section{Displacements exclusion in configuration of two double pumped Fabry-Perot cavities}\label{sec_2FP_cavities}

Comparing formulas (\ref{Cph}) and (\ref{barCph}) we see that platform displacements ($u_x$ and $u_y$) make different contributions. It allows to exclude, for example, displacement $y$ ($u_y$) in the following combination:
\begin{align}
 C_1&=\sqrt \frac{T_1R_2}{T_2R1}\, \frac{1+\psi^2}{2}\, C_{ph} -
        \sqrt \frac{T_2R_1}{T_1R_2}\, \psi\, \bar C_{ph}=\nonumber\\
 \label{C1}
 & =g_\text{1\,vac}-\\
 & -
        \frac{\sqrt 2 i\sqrt{T_1T_2R_1R_2} }{ \big(1-R_1R_2\psi^2\big)}
         \left(\left[\frac{1-\psi^2}{2}\right]^2u_x - \frac{\psi(1-\psi)^2}{2}\, u_h\right) ,\nonumber\\
&g_\text{1\,vac} \equiv \frac{\big(\psi^2-R_1R_2\big)\psi }{
    1-R_1R_2\psi^2}\left(
        \sqrt \frac{T_1R_2}{T_2R_1}\, \frac{1+\psi^2}{2}\,\frac{\big(b+ b_{-}^\dag\big)}{\sqrt 2}  -
        \right.\nonumber\\
 \label{g1vac}
 &\qquad -\left.       \sqrt \frac{T_2R_1}{T_1R_2}\, \psi\,
    \frac{\big(\bar b+\bar b_{-}^\dag\big)}{\sqrt 2}\right).
\end{align}
Here we denote by $g_\text{1\,vac}$ the linear combination of vacuum fluctuations $b$ and $\bar b$ incoming into cavity through non-pumped ports.

It is a very important result --- exclusion of information on $u_y$ is equivalent to conversion of platform $P_2$ into ideal mass, which is free from fluctuational displacement $y$.  The price for such conversion is decrease of GW response by factor approximately $\sim (1-\psi)^2$ (it is about $\sim(\Omega \tau)^2$ in long wave approximation) as compared with conventional laser GW detector.

\begin{figure}
%\begin{center}
\includegraphics[width=0.5\textwidth]{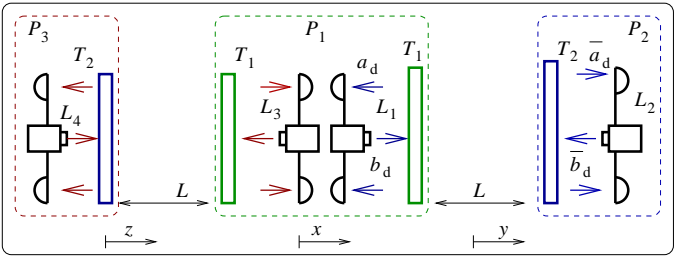}
%\end{center}
\caption{Configuration of two doubled pumped Fabry-Perot cavities. The right Fabry-Perot cavity is the same as shown on Fig,~\ref{T1T2oneFP}, the redirecting mirrors are not shown. The left Fabry-Perot cavity is identical to right cavity having the  mirror with transparency $T_1$ rigidly mounted on platform $P_1$. Left cavity is pumped by lasers $L_3$ and $L_4$, redirecting mirrors  are also not shown. }
\label{T1T2threeP}
\end{figure}

Now we have to exclude information on $u_x$ (i.e. displacement $x$ of platform $P_1$). It can be done in configuration of two double pumped Fabry-Perot cavities. Let us add  second Fabry-Perot cavity (left cavity on Fig.~\ref{T1T2threeP}) positioned in line with first cavity considered above. For simplicity we assume that parameters of both cavities are identical and that amplitudes and detunings of lasers $L_3,\ L_4$ pumped second cavity are the same as of lasers $L_1,\ L_2$ correspondingly.  Due to place shortage on Fig.~\ref{T1T2threeP} we could not show redirected mirrors assuming that complete scheme for each Fabry-Perot cavity is the same as shown on Fig.~\ref{T1T2oneFP} for one cavity. The mirrors with transparency $T_1$ and lasers $L_1,\ L_3$ with its detectors are rigidly mounted on the same platform $P_1$. The other mirror of second cavity and laser $L_4$ with its detectors are rigidly mounted on platform $P_3$, we denote its position  by $z$. We also assume that lasers $L_3$ and $L_4$ are tuned in resonance with second cavity and we measure phase quadrature components  by corresponding homodyne detectors.

In order to calculate formulas for phase quadratures of output waves $e_d^{(p)},\ \bar e_d^{(p)}$ of second cavity just rewriting formulas (\ref{adPh}, \ref{bdPh}) for phase quadratures $a_d^{(p)},\ \bar a_d^{(p)}$ we apply following substitutions:
\begin{subequations}
 \label{subs}
\begin{align}
 u_y & \to -u_z,\quad u_x \to -u_x,\quad u_h \to u_h, \\
    &b\to e, \quad \bar b\to \bar e.
\end{align}
\end{subequations}
Here amplitudes $e,\ \bar e$ describe corresponding vacuum noise incoming into second Fabry-Perot cavity though non-pumped ports.

The noise from lasers $L_3,\ L_4$ we exclude by the same manner as for first cavity. We can also exclude information on displacement $z$ in combination $C_2$ by the same way as we excluded displacement $y$ in combination $C_1$. One can write this combination $C_2$ free from displacement $z$ using substitutions (\ref{subs}):
\begin{align}
\label{C2}
 C_2&= g_\text{2\,vac}-\\
 & -   \frac{\sqrt 2 i\sqrt{T_1T_2R_1R_2} }{ \big(1-R_1R_2\psi^2\big)}
         \left(-\left[\frac{1-\psi^2}{2}\right]^2u_x - \frac{\psi(1-\psi)^2}{2}\, u_h\right) .\nonumber
\end{align}
Here $g_\text{2\,vac}$ is the combinations of vacuum noise amplitudes $e,\ \bar e$, described by the same formula (\ref{g1vac}) with only substitutions $b\to e, \quad \bar b\to \bar e$.

Comparing (\ref{C1}, \ref{C2}) we see that value $u_x$ makes contributions into $C_1$ and $C_2$ with the opposite signs, whereas GW contributions (i.e. $u_h$) have the same sign (it is obvious consequence of tidal nature of GW). So in order to exclude $u_x$ we should just sum $C_1$ and $C_2$:
\begin{align}
 C_\text{DFI} &= \frac{C_1 + C_2}{\sqrt 2} = \frac{g_\text{1\,vac} +g_\text{2\,vac}}{\sqrt 2} + \nonumber\\
 \label{CDFI}
 & \quad + \frac{i\sqrt{T_1T_2R_1R_2}}{
    \big(1-R_1R_2\psi^2\big)}
    \left( \psi\,\big(1-\psi\big)^2\, u_h\right) .
\end{align}
Comparing combination $C_\text{DFI}$ with combination $C_{ph}$ (\ref{Cph}) we see that gravitational signal in $C_\text{DFI}$ is smaller by factor $(1-\psi)^2$ which in  approximation of long gravitational wave length $L/\lambda_\text{grav}\ll 1$ (or $\Omega\tau\ll 1$) is about $(\Omega\tau)^2$.
It is the same decrease of GW response as in combination  $\tilde C_3$ (\ref{tildeC3}, \ref{tildeC3app})) for simplified model considered in Sec.~\ref{simple} (the only difference is the presence of resonance gain in (\ref{CDFI})).

Assuming  $T_1,\ T_2\ll 1$ and $\Omega\tau\ll 1$ we rewrite $C_\text{DFI}$ in narrow band  approximation:
\begin{align}
 \label{CDFIapp}
 C_\text{DFI} &\simeq \frac{g_\text{1\,vac} +g_\text{2\,vac}}{\sqrt 2}+
  \frac{\sqrt{T_1^3T_2}}{T_1^2+T_2^2}\,
    \frac{ \Omega^2\tau}{\gamma-i\Omega}\,A\, kL\, h(\Omega),
\end{align}
where $\gamma=(T_1^2+T_2^2)/4\tau$ is the relaxation rate (half bandwidth) of Fabry-Perot cavity.

Recall that in a simplest detector with two test masses and only one round trip of light between them  gravitational signal is about $AkLh$ with the same value of fluctuational field. So assuming in (\ref{CDFIapp}) that $\gamma\approx \Omega$ and $T_1\approx T_2$ we see that signal-to-noise ratio of our cavities operating as displacement noise free detector  is smaller by factor about $\sim \Omega\tau$ as compared with simplest detector.

\section{Conclusion}\label{conclusion}

In this paper we have analyzed the operation of two Fabry-Perot cavities positioned in line, performing the displacement-noise-free gravitational-wave detection. We have demonstrated that it is possible to construct a linear combination of four response signals which cancels {\em  displacement fluctuations} of the mirrors. At low frequencies the GW response of our cavities turns out to be better than that of the Mach-Zehnder-based DFIs \cite{2006_interferometers_DNF_GW_detection} due to the different mechanisms of noise cancellation.

Due to reflected and transmitted waves carry the same information on mirrors displacement we have additional possibility to exclude {\em laser  noise} (of course, fundamental vacuum noise can not be not excluded).

We show that considered DFI with two Fabry-Perot cavities is similar to the simplest round trip configuration shown in Fig.~\ref{DL3}.

For simplicity we have analyzed three platform configuration. The configurations with larger number of movable platform is more realistic and it may provide better sensitivity. For example, the middle platform may be splitted into three platforms:  two platforms  with mirrors (having transparency $T_1$) and one platform between them (with lasers $L_1,\ L_3$ and its detectors). Variants of such configurations are under investigation now.

The proposed configuration of DFI may be a promising candidate for the future generation of GW detectors  with  displacement  and laser noise exclusion which, in turn, will allow to overcome standard quantum limit.

\acknowledgments
We would like to thank V.B. Braginsky, Y. Chen, F.Ya. Khalili and S.P. Tarabrin for fruitful discussions. This work was supported by LIGO team from Caltech and in part by NSF and Caltech grant PHY-0651036 and by Grant of President of Russian Federation NS-5178.2006.2.

\appendix

\section{Derivation of formulas for Fabry Perot cavity}\label{derivation}

In this Appendix we derive formulas (\ref{ad}, \ref{bd}) for complex amplitudes  and (\ref{adPh}, \ref{bdPh}) for phase quadratures for single Fabry-Perot cavity pumped by laser from the left.

\begin{figure}
\includegraphics[width=0.5\textwidth]{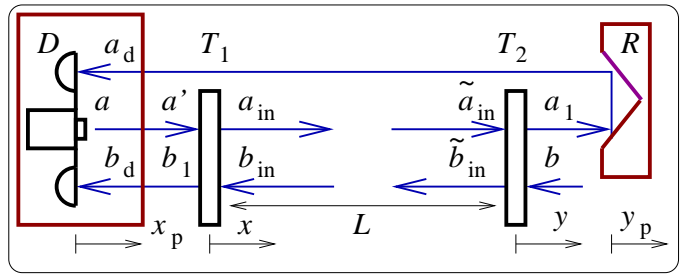}
\caption{Detailed scheme of measurement (generalization of shown on the Fig.~\ref{T1T2oneFP}a). Cavity mirrors are movable, laser and detectors are placed on detecting platform D, additional
 mirror is placed on reflecting platform $R$.}\label{fpthroughT1T2fourP}
\end{figure}

For methodical purpose we start from general case when laser with detectors, mirrors and additional mirror are mounted on separated rigid movable platform each as shown on Fig.~\ref{fpthroughT1T2fourP}. Below we use notations on Fig.~\ref{fpthroughT1T2fourP}. First we find complex mean amplitudes, writing boundary conditions on  right and left mirror:
\begin{align*}
\tilde B_{in} &= -R_2\tilde A_{in}, \quad A_1 = i\tilde A_{in} T_2,\\
A_{in}&=  iT_1\, A-R_1\, B_{in},\quad B_1=iT_1\,B_{in}-R_1\,A
\end{align*}
From these equations and obvious relations
$\tilde A_{in} = A_{in} \vartheta_0$ and $B_{in} = \tilde B_{in}  \vartheta_0$ one can find formula (\ref{Ain}) for $A_{in}$ and for mean output fields:
\begin{align}
\label{B1A2}
& B_1=\frac{A\big(R_2\vartheta_0^2-R_1\big)}{1-R_1R_2\vartheta_0^2},\quad
&  A_1 =  \frac {-T_1T_2\vartheta_0\, A}{1 -R_1R_2\vartheta_0^2}.
\end{align}

To find small amplitudes inside cavity we write down boundary condition on right and left mirrors correspondingly:
\begin{align}
\label{tildebin}
\tilde b_{in}&=    -R_2\tilde a_{in} + iT_2\,b -R_2\vartheta_0u_y\\
\label{ainA}
a_{in}&=iT_1\, a' - R_1b_{in} -R_1R_2\vartheta_0^2u_x.
\end{align}
And taking into account GW action as effective variation of refractive index $1+h/2$
\begin{align}
\tilde a_{in}&=
    \vartheta_0\psi a_{in} + A_{in}\vartheta_0\, i\,j(\Omega),\\
b_{in} &=
    \vartheta_0\psi \tilde b_{in} -R_2 A_{in}\vartheta_0^2\,i\,j(\Omega),\\
j&=\frac{\omega_0h(\Omega)}{2} \int_{t-\tau}^t e^{-i\Omega (t'-t)}\, dt'=
   \frac{kLh}{2}\left(\frac{1-e^{i\Omega\tau}}{-i\Omega\tau}\right), \nonumber\\
&\psi u_h = A_{in}\big(1+\psi\big) j\, \nonumber
\end{align}
we find small amplitudes inside cavity:
\begin{align}
\label{ainFin}
a_{in}&= \frac{iT_1\, a' -iT_2R_1\vartheta_0\psi b}{1-R_1R_2\vartheta^2\psi^2} +
   \frac{R_1R_2\vartheta_0^2\big(\psi (u_y+u_h)-u_x\big)}{1-R_1R_2\vartheta^2\psi^2},\\
\label{binFin}
b_{in} &=\frac{-iT_1R_2\vartheta_0^2\psi^2\, a' +iT_2\vartheta_0\psi b}{1-R_1R_2\vartheta_0^2\psi^2}+\\          &\qquad + \frac{R_2 \vartheta_0^2\psi\big(
   R_1R_2\vartheta_0^2\psi u_x - u_y-u_h\big)}{1-R_1R_2\vartheta_0^2\psi^2}.\nonumber
\end{align}

Now using second boundary condition on right  mirror we can find transmitted wave $a_1$:
\begin{align}
a_1 &= -R_2\, b +iT_2\, \tilde a_{in} =   \nonumber\\
\label{a1}
= & {\cal R}_2 b +{\cal T}a'
   - \frac{iR_1R_2T_2\vartheta_0^3\psi}{1-R_1R_2\vartheta_0^2\psi^2}\,
   \left( u_x-\psi (u_y +u_h)\right).
\end{align}

By the same manner from second boundary condition on left mirror we find reflected wave $b_1$
\begin{align}
b_1 &= -R_1\, a' +iT_1\, b_{in} +(-R_1 A)\, 2ikx=\nonumber \\
\label{b1}
&={\cal R}_1 a'+ {\cal T}b + \frac{i(R_1-R_2\vartheta_0^2)}{T_1}\, u_x+\\
& \qquad + \frac{iT_1R_2\vartheta_0^2}{1-R_1R_2\vartheta_0^2\psi^2}
   \big(  u_x- \psi (u_y+u_h)\big).\nonumber
\end{align}
Here we write formula for $b_1$ in this form in order to extract term  proportional to same combinations of  mirrors positions as in (\ref{a1}).

In order to express fields $a_1, \ b_1$ through small amplitude $a$ describing laser fluctuations we should substitute  in (\ref{a1}, \ref{b1})
\begin{align}
 a'& = \big(a -Aikx_p\big).
\end{align}
Now we can find field $b_d$ falling on detector
\begin{align}
 b_d &=b_1 - B_1\, ikx_p=
        {\cal R}_1 a+ {\cal T} b +
        \frac{i(R_1-R_2\vartheta_0^2)}{T_1}\, u_x+ \nonumber\\
 \label{bdfin}
 &\quad +       \frac{iT_1R_2\vartheta_0}{1-R_1R_2\vartheta^2}
   \big( \vartheta_0 u_x- \vartheta (u_y+u_h)\big) -\\
  -& \frac{A_{in} \big(1-R_1R_2\vartheta_0^2\big) ikx_p }{iT_1}\left(
    \frac{R_2\vartheta^2-R_1}{1-R_1R_2\vartheta^2}+
    \frac{R_2\vartheta_0^2-R_1}{1-R_1R_2\vartheta_0^2}\right).\nonumber
\end{align}
 
Using (\ref{a1}) for transmitted wave $a_1$ we find formula for amplitude $a_d$ falling on detector:
\begin{align}
 a_d &= \vartheta \big(-a_1 -A_1\, 2iky_p\big) +
    (-\vartheta_0 A_1) ij - (-\vartheta_0 A_1) ikx_p=\nonumber\\
&= -\vartheta_0\psi \big({\cal T} a +{\cal R}_2b\big)
    + \frac{iT_2R_1R_2\vartheta_0^4\psi^2
        \big(u_x-\psi (u_y+u_h)\big)}{1-R_1R_2\vartheta_0^2\psi^2}-\nonumber\\
\label{adfin}
&\qquad    - iT_2\vartheta_0^2\psi \big(u_h +A_{in}\, 2iky_p\big)+\\
&\qquad +     iT_2 \vartheta_0^2A_{in}ikx_p\left(1 +\psi^2
         \frac{\big(1-R_1R_2\vartheta_0^2\big)}{\big(1-R_1R_2\vartheta_0^2\psi^2\big)}
          \right).\nonumber
\end{align}
Now substituting $x_p=x$ and $y_p=y$ into (\ref{bdfin}, \ref{adfin}) one can obtain formulas (\ref{ad}, \ref{bd}).

It is useful to rewrite formulas (\ref{bdfin}, \ref{adfin}) for particular case of zero detuning  ($\vartheta_0=1$) and $x_p=x$ and $y_p=y$:
\begin{align}
\label{adFin}
 a_d &=  \frac{T_1T_2\psi^2 a - \psi(R_1\psi^2-R_2)b_2}{1-R_1R_2\psi^2}+\\
 &\quad    + \frac{iT_2 }{1-R_1R_2\psi^2}\left(\frac{1+\psi^2}{2}\, u_x -
    \psi(u_y+u_h)\right),\nonumber\\
\label{bdFin}
b_d &= \frac{(R_2\psi^2-R_1)\,a -T_1T_2\psi \, b_2}{1-R_1R_2\psi^2}+\\
 & \quad    \frac{iT_1R_2}{1-R_1R_2\psi^2}\left(
        \frac{1+\psi^2}{2}\, u_x-\psi\, \big(u_y+u_h\big)\right).\nonumber
\end{align}

We define phase quadratures of fields falling on detectors as fallowing:
\begin{align*}
a_d^{(p)} & \equiv\frac{-A_d^*a_d+A_da_{d-}^\dag }{i|A_d|\sqrt 2}=
    \frac{(\vartheta_0^*)^2a_d + \vartheta_0^2 a_{d-}^\dag }{ \sqrt 2},\\
b_d^{(p)} & \equiv \frac{B_d^*b_d-B_db_{d-}^\dag }{i|B_d|\sqrt 2}=
    \frac{\big(R_2(\vartheta_0^*)^2-R_1\big)b_d +
        \big(R_2\vartheta_0^2-R_1\big)b^\dag _{d-}}{|R_2\vartheta_0^2-R_1|\sqrt 2}\, .
\end{align*}
Substituting (\ref{adFin}, \ref{bdFin}) into these formulas we finally obtain formulas (\ref{adPh}, \ref{bdPh}) for phase quadratures $a_d^{(p)},\ b_d^{(p)}$.

%\bibliography{noise_free}

\end{document}